\documentclass[11pt]{amsart}
\usepackage{latexsym,amssymb,amsmath,amscd,amsthm}\usepackage{amsfonts}
\numberwithin{equation}{section}
\theoremstyle{remark}

\newcommand{\bq}{\begin{equation}}
\newcommand{\bea}{\begin{array}}
\newcommand{\eea}{\end{array}}

\newcommand{\gD}{\Delta}

\newcommand{\gb}{\beta}

\newcommand{\mf}{\mathfrak}

\newcommand{\mc}{\mathcal}

\newcommand{\dg}{\dagger}

\newcommand{\ol}{\overline}

\newcommand{\gG}{\Gamma}

\newcommand{\gd}{\delta}
\newcommand{\pp}{\partial}

\newcommand{\na}{\nabla}

\newcommand{{\DDD}}{D\!\!\!\!\!\!-}

\title{THERMODYNAMICS AND SCALE RELATIVITY}

\author{Robert Carroll\\University of Illinois, Urbana, IL 61801}

\date{October, 2011\thanks{email: rcarroll@math.uiuc.edu}}

\begin{document}


\bibliographystyle{plain}


\begin{abstract} 
It is shown how the fractal paths of SR = scale relativity (following Nottale) can be introduced into a TD = thermodynamic context (following Asadov-Kechkin).
\end{abstract}

\maketitle


\section{PRELIMINARY REMARKS}
\renewcommand{\theequation}{1.\arabic{equation}}
\setcounter{equation}{0}

The SR program of Nottale et al (cf. \cite{notl} has produced
a marvelous structure for describing quantum phenomena on
the QM type paths of Hausdorff dimension two (see below).
Due to a standard Hamiltonian TD dictionary (cf. \cite{ptrn})
an extension to TD phenomena seems plausible.  However
among the  various extensive and intensive variables of TD
it seems unclear which to embelish with fractality.  We avoid
this feature by going to \cite{askn} which describes the arrow of time
in connection with QM and gravity.  This introduces a complex
time $({\bf 1A})\,\,\tau=t-(i\hbar/2)\gb$ where $\gb=1/kT$ with
$k=k_B$ the Bolzmann constant and a complex Hamiltonian
$({\bf 1B})\,\,{\mf H}={\mf E}-(i\hbar\gG/2)$ where ${\mf E}$ is a standard energy term, e.g. 
$({\bf 1C})\,\,{\mf E}\sim (1/2)mv^2+{\mf W}(x)$.  One recalls that complex
time has appeared frequently in mathematical physics.
We will show how 
fractality can be introduced into the equations of \cite{askn}
without resorting to several complex variables or quaternions.
\\[3mm]\indent
Thus from \cite{askn} one has equations
\bq\label{1.1}
{\mf H}={\mf E}-\left(\frac{i\hbar}{2}\gG\right);\,\,\tau=t-\frac{i\hbar}{2}\gb;\,\,
[{\mf E},\gG]=[{\mf H},{\mf H}^{\dg}]=0;
\end{equation}
$$\Psi=exp^{-\frac{i{\mf H}\tau}{\hbar}}\psi;\,\,P_n=\frac{w_n}{Z};\,\,w_n=\rho_nexp^{[-{\mf E}_n
\gb+\gG_nt]};$$
$$i\hbar\pp_{\tau}\Psi={\mf H}\Psi;\,\,\Psi=\sum C_n\psi_n;\,\,
{\mf H}_n={\mf E}_n-\frac{i\hbar}{2}\gG_n;\,\,[{\mf H},{\mf H}^{\dg}]=0;$$
$${\mf E}\psi_n={\mf E}_n\psi_n;\,\,\gG\psi_n=\gG_n\psi_n;\,\,(\psi_n,\psi_k)=\gd_{nk}$$
One could introduce another complex variable here, say $j$ with $j^2=-1$, but this can be avoided.  
\\[3mm]\indent
Now go to the SR theory and recall the equations
\bq\label{1.2}
\frac{\hat{d}}{dt}=\frac{1}{2}\left(\frac{d_{+}}{dt}+\frac{d_{-}}{dt}
\right)-\frac{i}{2}\left(\frac{d_{+}}{dt}-\frac{d_{-}}{dt}\right);\,\,
{\mc V}=\frac{\hat{d}x}{dt}=V-iU=
\end{equation}
$$=\frac{1}{2}(v_{+}+v_{-})-\frac{i}{2}(v_{+}-v_{-});\,\,\frac{\hat{d}}{dt}=\pp_t+{\mc V}\cdot\na-i{\mc D}\gD$$
\bq\label{1.3}
{\mc H}=\frac{m}{2}{\mc V}^2-im{\mc D}\na\cdot{\mc V}+{\mc W}=\frac{1}{2m}{\mc P}^2-i{\mc D}\cdot{\mc P}+{\mc W};
\end{equation}
$${\mc H}={\mc V}\cdot{\mc P}-i{\mc D}\na\cdot{\mc P}-
{\mc L}$$
\bq\label{1.4}
\hat{{\mc V}}={\mc V}-i{\mc D}\na;\,\,(\pp_t+\hat{{\mc V}}\cdot\na){\mc V}=-\frac{\na{\mc W}}{m}
\end{equation}
\bq\label{1.5}
U={\mc D}\na log(P);\,\,P=|\psi|^2;\,\,\psi=e^{i{\mf S}/2m{\mc D}};\,\,{\mf Q}=-2m{\mc D}^2\frac{\gD\sqrt{P}}{\sqrt{P}};
\end{equation}
\bq\label{1.6}
{\mc V}=-2i{\mc D}\na[log(\psi)];\,\,{\mf S}_0=2m{\mc D};\,\,
{\mc D}^2\gD\psi+i{\mc D}\pp_t\psi-\frac{{\mc W}}{2m}\psi=0
\end{equation}
$$\frac{dV}{dt}=\frac{F}{m}=U\cdot\na U+{\mc D}\gD U$$
This has been written for 3 space dimensions but we will
restrict attention to a 1-D space based on $x$ below.
\\[3mm]\indent
We will combine the ideas in (1.1) and (1.2) in Section 2 below.
Note here ${\mf Q}$ is the QP= quantum potential (see e.g.
\cite{c006,c007,c009,clrr} for background).

\section{COMBINATION AND INTERACTION}
\renewcommand{\theequation}{2.\arabic{equation}}
\setcounter{equation}{0}

From (1.2)-(1.6) we see that the fractal paths in one space dimension
have Hausdorff dimension 2 and we note that U in (1.2)
is related to an osmotic velocity and completely determines the QP ${\mf Q}$.  Note that these equations (1.2)-(1.6)
produce a macro-quantum mechanics (where ${\mc D}=\hbar/2m$ for QM).  It is known that a QP represents a stabilizing organizational anti-diffusion force which suggests an important 
connection between the fractal picture above and biological
processes involving life (cf. \cite{aunl,notl,rpgo,sahz,zak,zaak}.  We also
refer to \cite{cath,naga,nels} for probabalistic aspects of quantum mechanics and entropy and recommend a number
of papers of Agop et al (cf. \cite{agop}) which deal with fractality (usually involving Hausdorff
dimension 2 or 3) in differential equations such as Ginzburg-Landau, Korteweg de-Vries, and Navier-Stokes; this work includes some formulations in Weyl-Dirac geometry (Feoli-Gregorash-Papini-Wood formulation) involving superconductivity in a gravitational
context.
\\[3mm]\indent
Now let us imagine that ${\mc W}\sim{\mf W}$ and $V\sim v$
so that the energy terms in the real part of the SE arising 
from (1.2)-(1.6) will take the form
\bq\label{2.1}
{\mf E}\sim \frac{1}{2}mV^2+{\mf W}+{\mf Q}
\end{equation}
and we identify this with ${\mf E}$ in the TD problem where
\bq\label{2.2}
{\mf Q}=-2m{\mc D}^2\frac{\gD\sqrt{P}}{\sqrt{P}};\,\,P=|\Psi|^2
\end{equation}
One arrives at QM for ${\mc D}=\hbar/2m$ as mentioned above
and one notes that the mean value $\bar{{\mf E}		}$ used in the analysis of \cite{askn} will now have the form
\bq\label{2.3}
\bar{{\mf E}}=\frac{1}{2}\int mV^2Pdx+\int |{\mf W}|^2Pdx+\int {\mf Q}Pdx
\end{equation} 
and the last term $\int {\mf Q}Pdx$ has a special meaning in terms of Fisher information as developed in \cite{c006,c007,c009,fred,frgy,fppo}.  In fact one has
\bq\label{2.4}
\int {\mf Q}Pdx=-2m{\mc D}^2\int\frac{\pp^2_x\sqrt{P}}{\sqrt{P}}
Pdx=
\end{equation}
$$=-\frac{{\mc D}^2}{2}\int\left[\frac{2P''}{P}-\left(\frac{P'}{P}\right)^2\right]Pdx=\frac{m{\mc D}^2}{2}\int\frac{(P')^2}{P}dx$$
In the quantum situation ${\mc D}=\hbar/2m$ leading to
\bq\label{2.5}
\int{\mf Q}Pdx=\frac{\hbar^2}{8m}\int \frac{(P')^2}{P}dx=\frac{\hbar^2}{8m}FI
\end{equation}
where $FI$ denotes Fisher information (cf. \cite{c009,fppo}).
and this term can be construed as a contribution from fractality.
\\[3mm]\indent
One can now sketch very briefly the treatment of \cite{askn} based on (1.1).
Thus one constructs a generalized QM (with arrow of time
and connections to gravity for which we refer to \cite{askn}).
The eigenvalues ${\mf E}_n,\,\,\gG_n,$ in (1.1)
are exploited with
\bq\label{2.6}
\rho_n=|C_n|^2;,\,P_n=\frac{w_n}{Z};\,\,\Psi=\sum C_n\psi_n;\,\,
w_n=\rho_ne^{-{\mf E}_n\gb+\gG_nt}
\end{equation}
One considers two special systems:
\begin{enumerate}
\item
First let the eigenvectors $\gG_n$ all be the same (decay
free system) and then $w_n=\rho_nexp[-{\mf E}_n\gb]$ which means
that $\gb$ is actually the inverse absolute temperature (multiplied by $k_B$) when ${\mf E}_n$ is identified with the n-th
energy level and the system is decay free.
\item
Next let all the ${\mf E}_n$ be the same so $w_n=\rho_nexp[-\gG_nt]$ and all the $\gG_N$ have the sense of decay parameters
if $t$ is the conventional physical time.
\end{enumerate}
Thus the solution space of the theory space can be decomposed into the direct sum of subspaces which have a given value of the energy or of the decay parameter.
It is seen that for $\gb=constant$ the dynamical equation for the basis probabilities is
\bq\label{2.7}
\frac{dP_n}{dt}=-(\gG_n-\bar{\gG})P_n;\,\,\frac{d\bar{\gG}}{dt}
=-D_{\gG}^2;\,\,D_{\gG}^2=\ol{(\gG-\bar{\gG})^2}
\end{equation}
From (2.7) one sees that $\bar{\gG}(t)$ is not increasing
which means that the isothermal regime of evolution has an arrow of time, which is related to the average value of the decay
operator.  Thus $P_n$ increases if $\bar{\gG}>\gG_n$ and 
decreases when $\bar{\gG}<\gG_n$.  One can also show that in the general case of $\gb=\gb(t)$ the dynamical equations
for the $P_n$ have the form
\bq\label{2.8}
\frac{dP_n}{dt}=-\left[\gG_n-\bar{\gG}+({\mf E}_n-\bar{{\mf E}})\frac{d\gb}{dt}\right]P_n
\end{equation}
Here the specific function $d\gb/dt$ must be specified or extracted from a regime condition $f(t,\gb,\bar{{\mf A}}(t,\gb))=0$ 
for some observable ${\mf A}$ (e.g. $\bar{{\mf E}}=constant$ is an adiabatic condition).
In the adiabatic case for example when $\bar{{\mf E}}=\sum_n{\mf E}_nP_n=constant$ there results
\bq\label{2.9}
\frac{d\gb}{dt}=-\frac{\ol{{\mf E}T}-
\bar{{\mf E}}\bar{T}}{D^2_{\mf E}}
\end{equation}
where $D_{\mf E}$ denotes a dispersion of the energy operator ${\mf E}$.
Using (2.8)-(2.9) one obtains 
\bq\label{2.10}
\frac{d\bar{\gG}}{dt}=-D_{\gG}^2\left[1-\frac{(\ol{{\mf E}T}-
\bar{{\mf E}}\bar{T})^2}{D_{\mf E}^2D_{\gG}^2}\right]\geq 0
\end{equation}
Subsequently classical dynamics is considered for $\hbar\to 0$
and connections to gravity are indicated with kinematically
independent geometric and thermal times (cf. \cite{askn}).


\newpage
\begin{thebibliography}{cccc}



%
\bibitem{agop} M. Agop et al, Chaos, Solitons, and Fractals,
8 (1999), 1295; 16 (1999), 3367; 17 (2000), 3527, 16 (2003),
321; 30 (2006), 441; 32 (2006), 30; 34 (2007), 172; 38 (2008),
1243; Jour. Math. Phys., 46 (2005), 062110; Class. and Quant.
Gravity, 16 (1999), 3367, 17 (2000), 3627, 18 (2001), 4743;
Euro. Phys. Jour. D, 49 (2008), 35; 56 (2010), 239; Gen. Relativ.
and Grav., 40 (2008), 35
%
\bibitem{afis} D. Acosta, p. Fernandez de Cordoba, J. Isidro,
and J. Santandar, hep-th 1107.1898
%
\bibitem{askn} V. Asadov and O. Kechkin, hep-th 0608148,
0612122, 0612123, 0612162, 0702022, and 0702046; 
Moscow Univ. Phys. Bull., 63 (2008), 105-108; Grav. and
Cosmology, 15 (2009), 295-301
%
\bibitem{aunl} C. Auffray and L. Nottale, Progress in biophysics
and molecular biology, 97, 79 and 115
%
\bibitem{c006} R. Carroll, Fluctuations, information, gravity, and the quantum
potential, Springer, 2006
%
\bibitem{c007} R. Carroll, On the quantum potential, Arima
Publ., 2007
%
\bibitem{c009} R. Carroll, On the emergence theme of physics,
World Scientific, 2010
%
\bibitem{clrr} R. Carroll, math-ph 1007.4744; gr-qc 1010.1732
and 1104.0383
%
\bibitem{cath} A. Caticha, quant-ph 1005.2357, 1011.0723,
and 1011.9746
%
\bibitem{fred} B. Frieden, Physics from Fisher information,
Cambridge Univ. Press, 1998; Science from Fisher information,
Springer, 2004
%
\bibitem{frgy} B. Frieden and R. Gatenby, Exploratory data analysis using Fisher information, Springer, 2007
%
\bibitem{fppo} B. Frieden, A. Plastino, A.R. Plastino, and B. Soffer, Phys. Rev. E, 60 (1999), 48-53 and 60 (2002), 046128;
Phys. Lett. A, 304 (2002), 73-78
%
\bibitem{grpp} D. Gregorash and G. Papini, Nuovo Cimento,
B, 55 (1980), 37-51, 56 (1980), 21-38, and 63 (1981),
487-509
%
\bibitem{ljvz} G. Lebon, D. Jou, and J. Casas-Vazquez, Understanding non-equilibrium TD, Springer, 2008
%
\bibitem{naga} M. Nagasawa, Schr\"odinger equations and
diffusion theory, Birkh\"auser, 1993; Stochastic processes
in quantum physics, Birkh\"auser, 2000
%
\bibitem{nels} E. Nelson, Quantum fluctuations, Princeton
Univ. Press, 1985; Phys. Rev. 130 (1966), 1079;\,\,Dynamical
theory of Brownian motion, Princeton Univ. Press, 1967
%
\bibitem{notl} L. Nottale, 
Scale relativity and fractal space-time, Imperial College Press, 2011
%
\bibitem{ottg} H. Ottinger, Beyond equilibrium TD, Wiley, 2005;
Phys. Rev. E, 73 (2006), 036126
%
\bibitem{ptrn} M. Peterson, Amer. Jour. Phys., 47 (2979),
488-490
%
\bibitem{rpgo} R. Roman-Roldan, P. Bernaola-Galvan, and J.
Oliver, Pattern Recog., 7 (1187-1194),1996
%
\bibitem{sahz} I. Sanchez, Jour. Mod. Phys., 2 (2011), 4 pages,%
\bibitem{zak} M. Zak, Inter. Jour. Theor. Phys. (IJTP), 32 (1992), 159-190;
33 (1994), 2215-2280;
Chaos, Solitons, and Fractals, 9 (1998), 113-1116;
10 (1999), 1583-1620; 11 (2000), 2325-2390; 13 (2002), 39-41;  
32 (2007), 1154-1167; 2306; Phys. Lett. A, 133 (1989),
18-22 and 255 (1999),110-118;
Information Sciences, 128 (2000), 199-215 and 
129 (2000), 61-79; 165 (2004), 149-169
%
\bibitem{zaak} M. Zak, Int. Jour. Thero. Phys., 33 (1994), 1113-1116; 39 (2000), 2107-2140;
Chaos, Solitons, and Fractals, 14 (2002), 745-758; 19 (2004), 645-666; 26 (2005), 1019-1033, 28 (2006), 616-626; 32 (2007),1154-1167; 34 (2007), 344-352; 41 (2009),
1136-1149 and 2306-2312; 42 (2009), 306-315; Found. Phys.
Lett., 15 (2002), 229-243
%

\end{thebibliography}
\end{document}